\documentclass[twocolumn,showpacs,preprintnumbers,amsmath,amssymb,aps,pre]{revtex4}

\usepackage{mathtools,amssymb,lipsum}

\usepackage{epsfig}

\usepackage{graphicx,latexsym,xcolor}

\usepackage{tikz}

\usepackage[english]{babel}
\usepackage{amsmath}
\usepackage{color}
\usepackage{epsfig}
\usepackage{graphicx}
\usepackage{bm}
\usepackage{times,amsmath,amssymb}
\usepackage{subfigure}
\usepackage{amsfonts}
\usepackage{amssymb}
\usepackage{color}
\global\long\def\bege{\begin{equation}}
\global\long\def\ende{\end{equation}}

\global\long\def\begal{\begin{align}}
\global\long\def\endal{\end{align}}

\begin{document}

\title{The principle of least action for random graphs}
\author{Ioannis Kleftogiannis$^1$, Ilias Amanatidis$^{2}$}
\affiliation{$^1$ Physics Division, National Center for Theoretical Sciences, Hsinchu 30013, Taiwan }
\affiliation{$^2$Department of Physics, Ben-Gurion University of the Negev, Beer-Sheva 84105, Israel}

\date{\today}
\begin{abstract}
We study the statistical properties of the physical action S for random graphs, by treating the number of neighbors at each vertex of the graph (degree), as a scalar field. For each configuration (run) of the graph we calculate the Lagrangian of the degree field by using a lattice quantum field theory(LQFT) approach. Then the corresponding action is calculated by integrating the Lagrangian over all the vertices of the graph. We implement an evolution mechanism for the graph by removing one edge per a fundamental quantum of time, resulting in different evolution paths based on the run that is chosen at each evolution step. We calculate the action along each of these evolution paths, which allows us to calculate the probability distribution of S. We find that the distribution approaches the normal(Gaussian) form as the graph becomes denser, by adding more edges between its vertices. The maximum of the probability distribution of the action corresponds to graph configurations whose spacing between the values of S becomes zero $\Delta S=0$, corresponding to the least-action (Hamilton) principle, which gives the path that the physical system follows classically. In addition, we calculate the fluctuations(variance) of the degree field showing that the graph configurations corresponding to the maximum probability of S,
which follow the Hamilton's principle, have a balanced structure between regular and irregular graphs.

\end{abstract}

\maketitle
We investigate the physical properties of random systems that might lack a well defined spatial dimension or have one that dynamically fluctuates \cite{paper1,paper2}. Such spatio-dimensionless systems provide a good framework for studying fundamental problems such as the emergence of spacetime and its properties in emergent or quantum gravity approaches, using for example discrete models\cite{paper1,paper2,rovelli1,rovelli2,bombelli,fay1,fay2,surya,wolfram,gorard,markopoulou,trugenberger,trugenberger2}. Random graphs\cite{erdos_gallai,aigner,farkas,newman,frieze,berg,mizutaka,paper1,paper2} are good candidates for such studies, since they inherently posses a random geometry that can result in variable/fluctuating spatial dimensionality and many other interesting physical features\cite{paper1,paper2}. We explore the principle of least action (Hamilton's principle)\cite{dirac, feynman1,feynman2,infeld,schwinger} for such discrete random systems.

We consider the simplest case of random graphs, consisting of $n$ vertices, forming the set $V$, and $m$ edges randomly distributed among the vertices, forming the set $E$. Each edge connects only two vertices. There are $\Omega=\binom{ \binom{n}{2}}{m}$ possible configurations (runs) of the graph, all having the same probability to appear $P=1/\Omega$. These configurations form the ensemble
of the so-called uniform random graphs $G(V,E)$\cite{erdos_gallai,aigner,farkas,newman,frieze,berg,mizutaka,paper1,paper2}, containing the most generic and simple discrete random structures, with the minimum number of assumptions/restrictions. For ratio of edges over vertices
$R=\frac{m}{n}>0.5$ there is a giant(largest) component in the graph containing most of the vertices and edges.
The rest of the vertices and edges spread along many small disconnected components, which have mostly tree-like structures.
The number of these small components reduces as the number
of edges $m$ is increased.

\begin{figure}
\begin{center}
\includegraphics[width=0.9\columnwidth,clip=true]{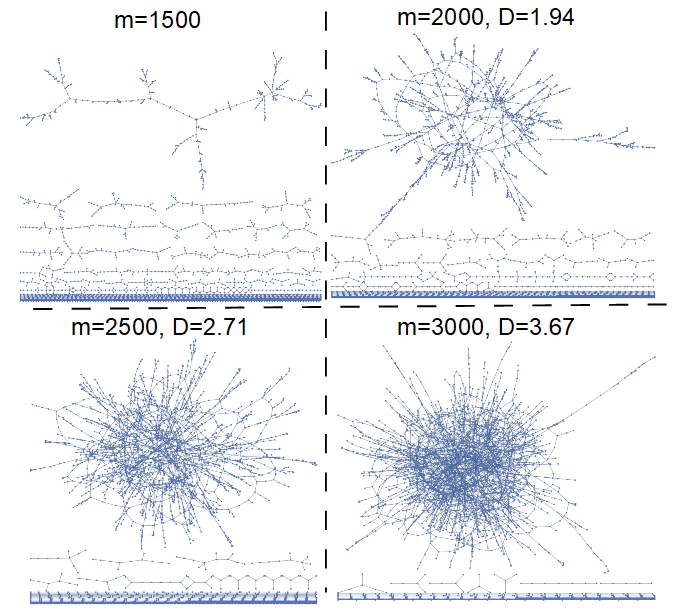}
\end{center}
\caption{Random graph structures for $n=3000$ number of vertices
and various numbers of edges $m$ distributed among the vertices.
In all cases a giant graph component appears containing most of the vertices and edges, along with many small disconnected components with tree-like structures. The emergent spatial dimension D of the giant component increases as the graph becomes denser with increasing m. For the sparser graph with m=1500 the giant component has a tree-like structure.}
\label{fig1}
\end{figure}

In Fig. \ref{fig1} we show one run of the graph for $n=3000$ and various $m\ge\frac{n}{2}$. In all cases the giant graph component appears along many small disconnected ones. For the rest of our study we focus on this giant component only. For the sparser graph with $m=1500$ the giant component has a tree-like structure. As more edges are added between the vertices in the graph, by increasing $m$, the giant component obtains a structure that has an emergent spatial dimension D, whose values are shown also in Fig. \ref{fig1}, calculated using a scaling(growth) method\cite{paper1,paper2}.
The spatial dimension is increased as the graph becomes denser by increasing $m$. In a previous study we have found that the giant component in a large graph, converges to a continuous 3-dimensional space(manifold) with flat geometry, at large scales and ratio $R \gtrsim 0.5$\cite{paper2}. This is important in the context of emergent spacetime from discrete models, showing that random graphs could be good candidates to model such mechanisms.

We consider that the number of neighbors at each vertex,
the degree $d(i)$ at vertex $v_i$, forms a scalar field
\begin{equation}
\phi_i=d(i)-2\frac{m}{n},
\label{field}
\end{equation}
spreading over the whole graph, where $\langle d(i) \rangle=2\frac{m}{n}=2R$ is the average degree over all the vertices for one run. The degree field Eq. \ref{field} has a real value at each vertex, which can be considered as a spatial point in the manifold that emerges when a large number of vertices are considered\cite{paper2}.
Since the number of edges for each run of the graph is fixed,
giving 
\begin{equation}
\sum_{i=1}^{n} d(i)=2m,
\label{field_con}
\end{equation}
the degree field satisfies the following constraint
\begin{equation}
\sum_{i=1}^{n} \phi_i= 0.
\label{field_con}
\end{equation}
The corresponding Lagrangian in a lattice-quantum-field-theory (LQFT) formulation for this degree field, that satisfies the U(1) symmetry, contains a term related to the potential energy of many coupled harmonic oscillators, which can be written at each vertex $v_i$ as,
\begin{equation}
\mathcal{L}_i = \sum_{j=1}^{n}  \phi_{i} L_{ij}\phi_{j},
\label{lagrangian}
\end{equation}
where \(L_{ij}\) is
the Laplacian matrix of the graph 
\begin{equation}
L_{ij}= D_{ij} - A_{ij},
\label{laplacian}
\end{equation}
corresponding to the Laplace operator in the continuous formulation
of QFT.
In the above equation \( D_{ij} \) is the degree matrix
\begin{equation}
D_{ij} =
\begin{cases}
d(i), & \text{if } i = j, \\
0, & \text{if } i \neq j.
\end{cases}
\end{equation}
and \( A_{ij} \) is the adjacency matrix
\begin{equation}
A_{ij} =
\begin{cases}
1, & \text{if } (v_i, v_j) \in E, \\
0, & \text{otherwise}.
\end{cases}
\end{equation}
The Lagrangian Eq. \ref{lagrangian} can be written as
\begin{equation}
\mathcal{L}_i=\phi_i^2 d(i) - \sum_{j=1}^n \phi_i A_{ij} \phi_j.
\end{equation}
The first term can be interpreted as a mass term for the degree field with effective mass $m_v=\sqrt{2d(i)}$.
We calculate the action for one configuration (run) of the graph by integrating the Lagrangian over all the vertices as
\begin{equation}
     S =  \sum_{i=1}^n \mathcal{L}_i. 
\label{action}
\end{equation}
The above formula can be expressed in terms
of the differences between the degree field values
of neighboring vertices, connected with an edge, as
\begin{equation}
     S =  \frac{1}{2} \sum_{i=1}^n \sum_{j=1}^n A_{ij}(\phi_i-\phi_j)^2, 
\label{action_dif}
\end{equation}
which is also known as the Dirichlet energy of the graph,
related to the uniformity of a function distributed on the graph, such as the degree field in our analysis. At each vertex $v_i$
we can define a local coordinate system of dimension d(i), with
each edge connecting the vertex to its neighboring ones, representing a spatial direction. From this point of view the field differences in Eq. \ref{action_dif} correspond essentially to the derivatives over different directions, in analogy to the Laplace operator in continuous physical systems.
From the above form Eq. \ref{action_dif} we can see that the average value of the degree used in the definition of the degree field $\phi_i=d(i)-2\frac{m}{n}$ does not play role in the action, since it is canceled by the field differences.
The average degree plays a role only if we include a mass
term in the field of the form $\frac{1}{2} (m_v\phi_{i})^2$,
by assigning for example a mass $m_v$ to each vertex in the graph. 

A complete graph, whose vertices are all connected
to each other has $m=\frac{n(n-1)}{2}$ edges in total.
For this case the degree is the same everywhere $d(i)=n-1$
and the degree field becomes uniform taking the constant value
\begin{equation}
\phi_i=\frac{n^2-n-2m}{n}.
\end{equation}
The sum in the Lagrangian Eq. \ref{lagrangian} gives a term
\begin{equation}
(n-1)(\frac{n^2-n-2m}{n})^2
\end{equation}
for i=j, and the opposite term for $i \neq j$, that cancel with each other, resulting in $\mathcal{L}_i=0$ and action $S=0$. This result is also true for regular graphs like the square or the cubic with periodic boundary conditions, that have the same degree everywhere $(d_{s}(i)=4,d_{c}(i)=6)$, or in the absence of edges in the graph (m=0,d(i)=0) when all the vertices are disconnected from each other. Therefore uniform degree fields result always in the trivial case S=0, since there is no potential energy contained in the degree field.

\begin{figure}
\begin{center}
\includegraphics[width=0.9\columnwidth,clip=true]{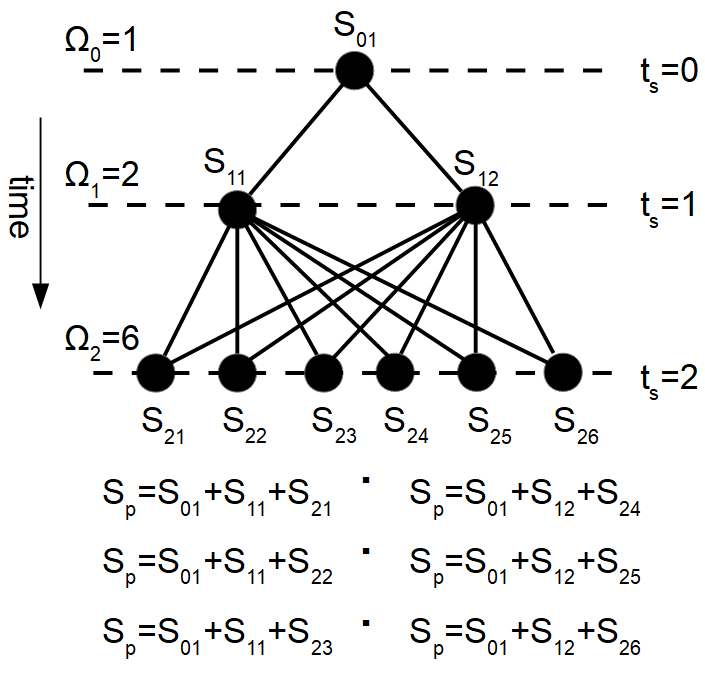}
\end{center}
\caption{The evolution mechanism of the graph represented
as a tree-like structure. Each node in the tree represents a graph configuration(run) $r_{t_s}$ at time step $t_s$, having a respective physical action value $S_{t_s,r_{t_s}}$. At each step $t_s$ one edge between the vertices of the graph is removed. The graph follows a path through the tree-like structure, by choosing randomly one run $r_{t_s}$ at each step $t_s$, with equal probability $P=1/\Omega_{t_s}$. Each path results in a different $S_p$ value determined by the sum over the values $S_{t_s,r_{t_s}}$ for the different runs the graph passes through, as it evolves through the tree-like structure.}
\label{fig2}
\end{figure}

We adopt an evolution mechanism for the graph by removing
one edge per a fundamental quantum of time $t_f$. The evolution mechanism is shown schematically in Fig. \ref{fig2}. At each step 
in the evolution, one of the possible configurations (runs) of the graph, determined by n and m, is chosen at random to represent the physical state of the system. Each run has the same probability $P=1/\Omega_{t_s}$ to appear, where $\Omega_{t_s}$ is
the number of runs at time step $t_s$, determined by n and m. As the system evolves it follows a path through the tree-like structure of possible runs. After $T$ evolution steps corresponding to time interval $(t=T t_f)$, the number of possible paths is
\begin{equation}
N_{p}(T)=\prod\limits_{t_s = 0}^{T} \Omega_{t_s}.
\end{equation}
All these paths have the same probability to appear 
\begin{equation}
P_{p}(T)=\frac{1}{N_p}.
\label{prob_path}
\end{equation}
Each path results in a different value of the action $S_p$,
which is found by summing over the action values $S_{t_s,r_{t_s}}$ of the runs across the evolution path
\begin{equation}
     S_p =  \sum_{t_{s}=0}^T S_{t_s,r_{t_s}},
\label{action_path}
\end{equation}
where $r_{ts} \in [1,\Omega_{t_s}]$ as shown in Fig. \ref{fig2}.
The sequence of the values of the index $r_{t_s}$ determines the evolution path. The statistical ensemble of the values given by Eq. \ref{action_path} allows us to study the statistical properties of the action $S_p$. 

\begin{figure}
\begin{center}
\includegraphics[width=0.9\columnwidth,clip=true]{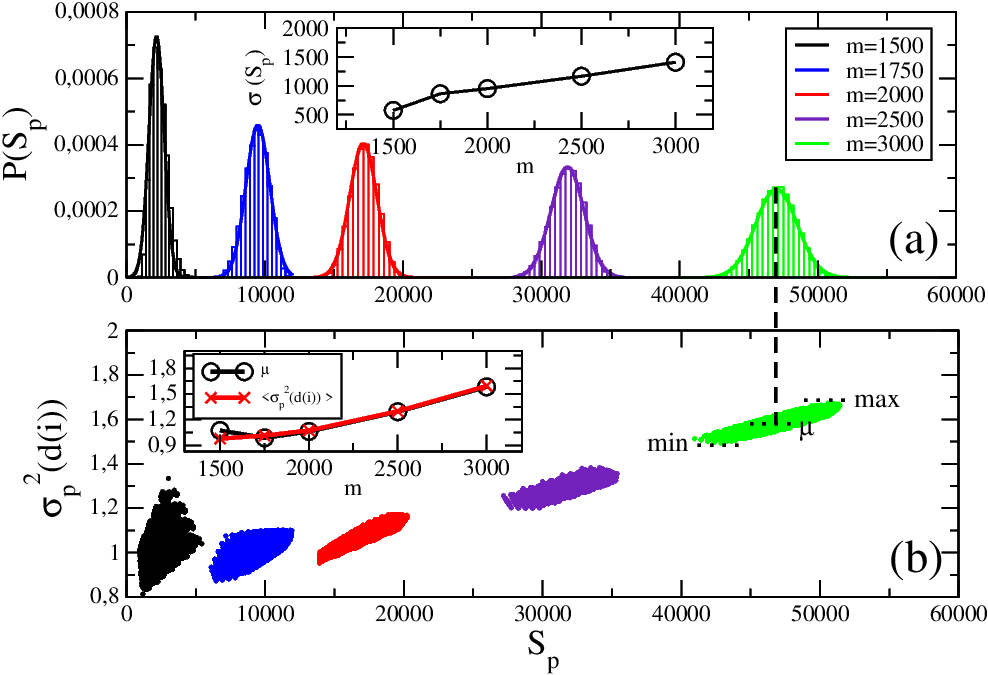}
\end{center}
\caption{a)The probability distributions of the action $S_p$ for
$n=3000$ number of vertices and $T=3$ time steps
in the graph evolution. We have considered 20 graph configurations (runs) at each evolution step. The different m correspond to the 
number of edges in the starting evolution step. Inset: The standard deviation of S. b) The variance of the number of neighbors (degree)
at each vertex in the graph. Inset: The medium value $\mu$ of the variance vs m, compared with the average variance for evolution paths that have the highest S degeneracy, corresponding to the principle of least action $\Delta S_p=0$, located at the maxima
of the distributions $P(S_p)$ in panel (a).}
\label{fig3}
\end{figure}

In Fig. \ref{fig3}a we plot the probability distribution of the action $P(S_p)$, for different number of edges $m$ in the starting runs of the graph evolution at time step $t_s=0$. For computational efficiency we have considered $n=3000$ number of vertices, a fixed number of runs at each evolution step $\Omega_{t_s}=20$, and $T=3$ steps, resulting in $N_p=\Omega_{t_s}^{T+1}=20^4$ evolution paths. In all cases the action is distributed normally as shown from the comparison between the histograms and the fitting Gaussian curves in Fig. \ref{fig3}a.
The fluctuations of $S_p$ determined by the distribution width, increase with increasing m, as the graph becomes denser. This effect can be seen also in the inset of Fig. \ref{fig3}a, where the standard deviation of $S_p$ is shown.
The action values increase in average with increasing number of edges $m$, since there are more differences in Eq. \ref{action_dif}. This effect will be reversed at some point when the graph starts becoming more uniform by approaching the complete graph structure. In this case all vertices are connected to each other, resulting in a uniform degree field and $S_p=0$. This is also true when there are no edges in the graph (m=0), and all vertices are disconnected from each other.

Classically based on the principle of least action (Hamilton's principle) we expect the system to follow evolution paths corresponding to $\Delta S_p=0$, where small deviations from the evolution path result in a minimal or no variation in the value of $S_p$. In general for a statistical physical system the differences $\Delta S_p$ between the values of $S_p$, become minimum at the maxima of the distributions $P(S_p)$, where the values of $S_p$ are more dense. For the graph we have found
that different evolution paths can result in the same action value $S_p$. This way different sets of evolution paths are formed, with each set containing paths giving the same $S_p$ value and therefore satisfying $\Delta S_p=0$. The number of paths in each set can be thought as the degeneracy of the set $\mathcal{D}(S_p)$. Then the probability of $S_p$ is simply
\begin{equation}
P(S_p)=\frac{\mathcal{D}(S_p)}{N_p}.
\label{prob_action1}
\end{equation}
The different sets of paths can be thought as different evolution histories of the physical system.
Classically we can assume that the system follows paths corresponding to the highest probability $P(S_p)$, 
at the maxima of the probability distributions
shown in Fig. \ref{fig3}a.
In addition, a quantum mechanical picture emerges,
if we consider that the system
samples all the possible paths, with equal probability given by Eq. \ref{prob_path}, across its evolution,
by assigning also a phase factor $e^{i \hbar S_p}$ to the wavefunction of each path\cite{dirac, feynman1,feynman2,infeld,schwinger}.

In Fig. \ref{fig3}b we plot the variance of the degree $\sigma_p^2(d(i))$ along the different evolution paths of the system. We have averaged over the variances $\sigma_{t_s,r_{t_s}}^2(d(i))$ of each run at time step $t_s$ along each evolution path in analogy to Eq. \ref{action_path} as, 
\begin{equation}
     \sigma_{p}^2(d(i)) =  \frac{1}{T+1}\sum_{t_{s}=0}^T \sigma_{t_s,r_{t_s}}^2(d(i)).
\label{variance}
\end{equation}
The variance allows us to estimate the amount of irregularity
in the graph structure. For example regular lattices, like
a square or a cubic lattice, have the same degree everywhere
$(d_{s}(i)=4,d_{c}(i)=6)$, resulting in $\sigma_p^2(d(i))=0$. For our graph model the values of $\sigma_p^2 (d(i))$ shown in Fig. \ref{fig3}b spread along small regions, due to the path degeneracy of $S_p$. In the inset of Fig. \ref{fig3}b we plot the medium value
\begin{equation}
   \mu=\sigma_p^2(d(i))_{min}+\frac{\sigma_p^2(d(i))_{max}-\sigma_p^2(d(i))_{min}}{2}, 
\end{equation}
along with the average value the $\langle \sigma_{p}^2(d(i)) \rangle$ over paths that have the highest $S_p$ degeneracy,
corresponding to the maximum probability $P(S_p)$ in Fig. \ref{fig3}a, described by
\begin{equation}
     \langle\sigma_{p}^2(d(i))\rangle =  \frac{1}{\mathcal{D}(S_p)}\sum_{p=1}^{\mathcal{D}(S_p)} \sigma_{p}^2(d(i)),
\end{equation}
where the sum runs over paths that have the highest $S_p$ degeneracy
$\mathcal{D}(S_p)$.
The two curves coincide showing that evolution paths corresponding to the most probable $S_p$, satisfying the principle of least action $\Delta S_p=0$, have a structure between regular graphs, corresponding to $\sigma_p^2(d(i))_{min}$, and irregular graphs, corresponding to $\sigma_p^2(d(i))_{max}$. Therefore the system evolves classically via balanced graph structures, between regular and irregular.

We note that adding a mass term for the degree field in Eq. \ref{lagrangian} of the form $\frac{1}{2} (m_v\phi_{i})^2$ does not qualitatively affects our result, shifting for example the probability distributions of $S_p$ without changing their Gaussian form.

To conclude we have shown how the least action principle
manifests in random graphs, by interpreting the degree at each vertex as a scalar field. We have shown that the statistics of the
physical action for the degree field follows the Gaussian
distribution. The graph evolves classically through configurations that have a balanced structure between regular and irregular graphs, corresponding to the most probable values of the action,
satisfying also the least action principle.
Quantum mechanically the graph samples different sets of paths,
with each set corresponding to the same action value.
Our results shows how physical scalar fields following the Hamilton's principle, can emerge from discrete models, such as those used to model spacetime in emergent and quantum gravity approaches. An interesting extension of the current study would be to investigate the emergence of other types of fields
different than scalar, with internal group symmetries,
such as those appearing in the standard model of particle physics.
More generally, further investigations of the least action principle and the relevant properties of the physical action
for other types of graphs with different types
of random geometries and dimensionality
properties, would be also interesting.

\end{document}